# A New Automatic Change Detection Framework Based on Region Growing and Weighted Local Mutual Information: Analysis of Breast Tumor Response to Chemotherapy in Serial MR Images

Narges Norouzi, Reza Azmi, *Alzahra University*

Nooshin Noshiri, *The University of Winnipeg*

Robab Anbiaee, *Shahid Beheshti University of Medical Science*

**Abstract**— The automatic analysis of subtle changes between longitudinal MR images is an important task as it is still a challenging issue in scope of the breast medical image processing. In this paper we propose an effective automatic change detection framework composed of two phases since previously used methods have features with low distinctive power. First, in the preprocessing phase an intensity normalization method is suggested based on Hierarchical Histogram Matching (HHM) that is more robust to noise than previous methods. To eliminate undesirable changes and extract the regions containing significant changes the proposed Extraction Region of Changes (EROC) method is applied based on intensity distribution and Hill-Climbing algorithm. Second, in the detection phase a region growing-based approach is suggested to differentiate significant changes from unreal ones. Due to using proposed Weighted Local Mutual Information (WLMI) method to extract high level features and also utilizing the principle of the local consistency of changes, the proposed approach enjoys reasonable performance. The experimental results on both simulated and real longitudinal Breast MR Images confirm the effectiveness of the proposed framework. Also, this framework outperforms the human expert in some cases which can detect many lesion evolutions that are missed by expert.

—————————— ◆ ——————————

## 1 INTRODUCTION

BREAST cancer is one of the common malign diseases among women in Iran and many other parts of the world. Medical imaging plays a pivotal role in breast cancer care including detection, diagnosis, and treatment monitoring. Currently, mammography is the primary screening modality that is widely used to detect and diagnose breast cancer. Unfortunately, it has some limitations [1], [2]. 10% to 30% of breast cancers goes undetected by mammography as its positive predictive value is less than 35% [3]. Hence, the use of other imaging modality such as MRI is increasing [4] in combination with mammography, especially for women at high risk.

Also MR images have a pivotal role in treatment moni-toring [5], [6]. Neoadjuvant chemotherapy is a standard treatment for patients with locally advanced breast cancer; however, using a method that can accurately monitor the response of patients is a challenging task because manual labeling and direct comparison of huge amount of data are difficult, error prone and sometimes impossible. Acquisition parameters and patients' position may change during scanning, and global or local deformation of anatomical structures may result in insignificant changes. In addition, many of the changes are very subtle and invisible to human experts. Therefore, an automatic change analysis method that is capable of detecting significant changes in longitudinal MRI sequence is vital for medical diagnosis, follow-up and prognosis. Unfortunately, as a variety of MRI artifacts make many unimportant changes among images and soft tissue of the breast makes the image registration a complex task, automatic change detection remains a challenging issue in breast medical image processing. Some methods have been proposed to detect changes in MR images [7], [8], [9], [10], [11], [12], [13], [14], [15], [16] but many of them carried out in the field of brain images and a few studies have focused on

_______________

- *Narges Norouzi, Reza Azmi are affiliated with the Medical Image Processing Laboratory (MIPL), Faculty of Engineering and Technology, Alzahra University, Vanak Village St., Tehran, Iran. E-mail: n.norozi@alzahra.ac.ir , azmi@alzahra.ac.ir.*
- *Nooshin Noshiri is affiliated with Terrabyte Research group, Department of Applied Computer Science, The University of Winnipeg, MB, Canada.*
- *Robab Anbiaee is affiliated the Department of Radiotherapy and Oncology, Shahid Beheshti University of Medical Science, Velenjak St., Shahid Chamran highway, Tehran, Iran. E-mail: Anbiaee@gmail.com.*

breast cancer monitoring. Since the structure and tissue of breast are very different from brain, we cannot use these methods directly. Also, these methods have some weaknesses that will be discussed.

Recent studies about the evaluation of breast lesions response to chemotherapy show that the most frequently applied methods include image registration followed by simple deterministic image subtraction [17], [18], [19], [20], [21] and tracking changes during the course of treatment in pharmacokinetic parameter values obtained from the tumor ROI or histograms describing their distributions [22], [23], [24]. However, all of these methods are limited. The first approach is very sensitive to noise due to direct intensity comparison. The second does not capture the tissue heterogeneity as observed in distributions of parameter values. Also, two last approaches discard the information on spatial localization and therefore suffer from a similar limitation as the RECIST criteria [18].

In this paper, we propose an automatic change detection framework that is composed of two phases, pre-processing and change detection. In each of the phases, new approaches have been presented to eliminate weak points of the previous studies. In the pre-processing phase, a method is suggested to extract regions containing changes which none of the previous methods had been proposed. This method can remove lots of pixels containing noise and unimportant changes to avoid reduction of change detection system performance.

In the second phase, by means of presented Weighted Local Mutual Information (WLMI) method, high level features are extracted which can better differentiate pixels containing significant changes from unreal ones compared to the intensity average and subtraction measure. Finally, the mentioned features are used by a region growing based algorithm to make final decision about detected changes. This algorithm eliminates some forged changes from final results due to adherence to the principle of local consistency of changes.

The paper is organized as follows. An overview of the related work is given in section 2. The two main phases of proposed framework are presented in section 3. Section 4 deals with the results obtained on both simulated and real images. The proposed approach for simulating images is first described in section 4.1 and results of the experiments on these simulated and real data are summarized in section 4.2 and 4.3 respectively. A discussion is provided in section 5 and conclusion is finally given in section 6.

## 2 RELATED WORKS

Recently, detecting significant changes in serial images of the same scene taken at different times has been getting a lot of attention for the large number of applications such as surveillance, remote sensing, medical diagnosis and civil infrastructure.

Some different methods have been proposed for change detection in serial medical images. But as mentioned in section 1, many of the given methods focused on brain disease monitoring such as cancer tumors [8], [11], [14], [15], [16], multiple sclerosis [9], [10], [12], Alz-heimer [28] and etc. Since a few of these methods have been applied to breast cancer monitoring, we review the related works from an overall perspective and point out their pros and cons. Based on comparison approaches that have been used in these methods, we can divide them into three categories, as you can see in Fig. 1. In the following we describe each comparison approach:

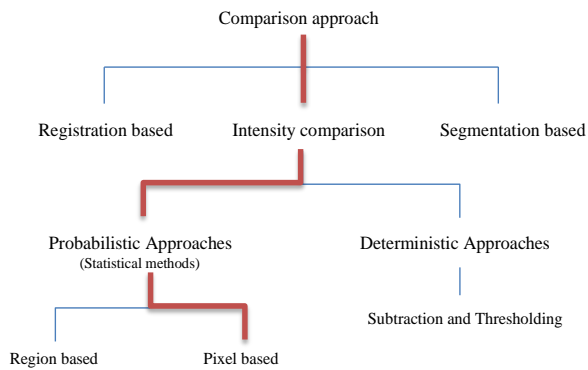

Fig. 1. Comparison approaches.

### Segmentation-Based Approach

In this approach, first the regions of interest have been segmented manually, semi-automatically, or fully automatically and then are compared to each other [25], [26], [27], [28]. In the recent years we have proposed several methods for medical image segmentation in different scopes such as brain tissue [29], [30] and breast tumor segmentation [31].

Segmentation-based change detection approach is a very simple method, but its performance highly depends on segmentation algorithms. Since the MR images are seriously affected by intensity inhomogeneities created by radio-frequency coils, accurate and automatic segmentation of lesions in these images remains a challenging task. Also large lesions may be sufficiently extracted, but distinguishing small lesions from noise and other structures is difficult [10].

### Registration-Based Approach

In the second category of comparison approaches, first an image registration method is used to obtain accurate geometrical alignment of images. Then aligned images are compared using simple deterministic image subtraction followed by manual or automatic thresholding [17], [18], [19], [20], [21]. Although most change detection systems [7], [9], [10], [12] use registration as a preliminary step, using only image subtraction as a comparison measure, makes these methods strongly sensitive to noise.

### Direct Intensity Comparison Approach

The last family of approaches relies on a direct comparison of intensities at the voxel level or on small regions [10]. Since in this approach intensities of pixels (or voxels) are compared directly, intensity normalization is a vital step to compensate global intensity changes. Generally direct comparison methods have two main stages: intensity normalization and comparison. The comparison of im-ages can be done using deterministic or probabilistic approach. Manual or automatic thresholding after simple image subtraction is a deterministic comparison that widely has been used in many of the studies [7], [8], [11], [12]. As emphasized in pervious section, using simple subtraction and thresholding decrease the performance of change detection system due to existence of noise in MR images.

Probabilistic methods build a statistical model of intensities and noise that is used to determine significant changes. Bosc et al. [10] proposed an automatic change detection system based on Generalized Likelihood Ratio Test (GLRT) that extended previous work [32] by handling multimodal image data with applications to follow-ups of multiple sclerosis. In addition, they developed a joint histogram-based technique for compensating nonlinear global intensity changes. Their method often fails when noise is nonstationary and cannot detect changes with low intensity variations (the lesions in early stages). Since the GLRT is a powerful statistical method to detect significant changes, Boisgontier et al. [9] recently proposed a parametric statistical testing framework which relies on the GLRT to detect changes between diffusion MRI acquisitions.

Patriarche and Erickson [14], [15] implemented an integrated change detection system based on post-classification of image's pixels in multispectral MR intensity feature space. They assumed that an abnormal tissue may look like a tissue transitioning from one normal tissue to another in the feature space. Clinical studies show that their system can identify subtle changes, but the tissue classification still remains very difficult and the whole process suffers from high time complexity [13].

Seo and Milanfar [16] proposed a non-parametric statistical approach that uses a single modality to find subtle changes and does not require prior knowledge of type of changes to be sought. However, their work does not address the effect of misalignment.

during the last decade are mostly statistical-based because of their power in detection of significant changes. Hence, the framework proposed here is based on a statistical approach that can eliminate some weak points of the previous studies. The selected approach is shown in red lines in Fig. 1.

We can summarize the strengths of our proposed framework as follows:
– Global intensity changes are removed due to using HHM.
– Local deformations are corrected owing to using a free-form deformation based on B-Splines functions.
– Lots of pixels containing noise and unimportant changes are removed in the beginning due to using proposed EROC method to extract regions of changes.
– Subtle changes are detected and a clear distinction is made between significant and unreal changes due to extracting high level WLMI features.
– Using region growing based algorithm which can eliminates unreal changes like noise from final results due to adherence to the principle of local consistency of changes.
– Finally, our proposed framework uses a single modality to find subtle changes and does not require prior knowledge about type of changes.

## 3 PROPOSED FRAMEWORK

As mentioned in introduction, the proposed framework composed of two main phases that its architecture is shown in Fig. 2. Each of these phases contains different stages which its subsystem is drawn separately (subsystem A and B) and they will be discussed in details.

### 3.1 Preprocessing Phase

This phase includes four stages which are as follows:

#### 3.1.1 Noise Reduction

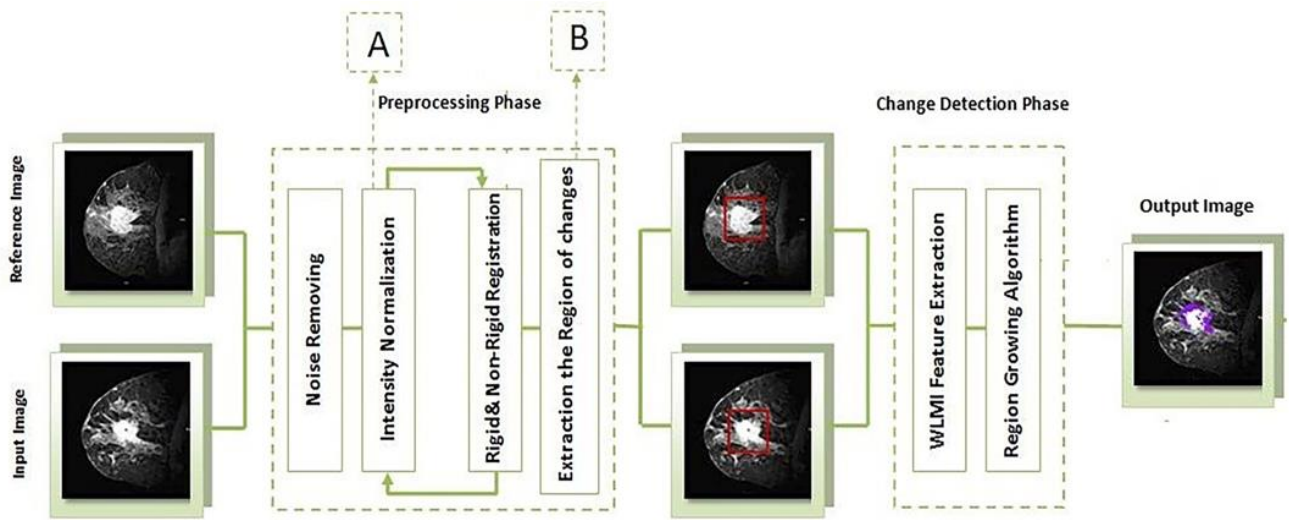

Fig. 2. Overall scheme of proposed change detection framework.

As described in this section, several studies conducted

According to the studies, noise in MR images contains

Rician distribution [33] and as this noise depends on signals and not Gaussian noise; therefore, separating it from signal is very hard. But fortunately Nowak [33] has presented a powerful method to remove Rician Noise from MR images which is based on wavelet filtering. In this research, the mentioned method is used to eliminate noise from images.

### 3.1.2. Intensity Normalization

The first stage in change detection methods based on direct comparison of images is removing inhomogeneity and global changes of intensity. These changes may occur because of the calibration errors and or variations in imaging system components. Many of the presented methods use linear intensity normalization functions to re-move this type of changes. But Bosc et al. [10] concluded that these changes are nonlinear so linear functions are not able to correct them. Therefore, they presented a method based on joint-histogram to estimate nonlinear normalization functions. Now, in this stage we have presented a method that enjoys less sensitivity towards noise due to using HHM. The overview of suggested method is shown in Fig. 3.

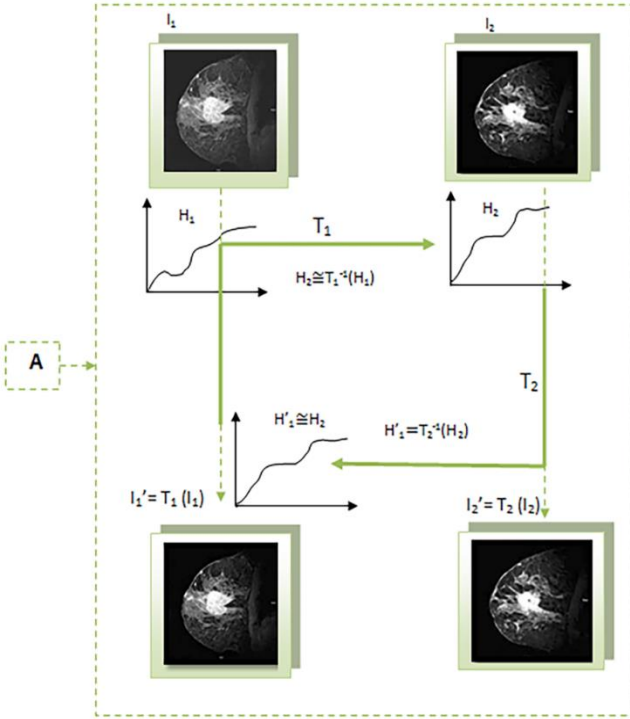

Fig. 3. HHM for intensity normalization.

Fig. 3 shows that the image histogram matching is used in each step for intensity normalization.

Let's consider images $I_1$ and $I_2$. In the first step by matching histogram $H_1$ to histogram $H_2$, transformation function $T_1$ is estimated based on intensity domain of image $I_2$ which is used to normalize the intensity of image $I_1$. The histogram of the achieved image $I_1'$ through transformation function $T_1$ is approximately equal to histogram $H_2$ that has been shown by $H_1'$ in (1).

$$H_2 \cong T_1^{-1}(H_1) \rightarrow I_1' = T_1(I_1), H_1' \cong H_2'$$ (1)

The second step deals with accurate intensity normalization. This time histogram $H_2$ is matched to histogram $H_1'$ to estimate transformation function $T_2$ which is used to normalize the intensity of $I_2$ based on $I_1'$.

$$H_1' \cong T_2^{-1}(H_2) \rightarrow I_2' = T_2(I_2), H_1' \cong H_2'$$

By using this method, intensity normalization can be performed with high accuracy.

Samples of images resulted from HHM method has been shown in Table 1.

### 3.1.3 Rigid and Non-Rigid Registration

The second and the most significant stage in preprocessing phase concerned with image registration, because of its effect on the final performance of the change detection system.

This stage involves correcting established deformation owing to different condition of image acquisition. In addition to transformation and rotation artifacts, as breast tissue is soft, its touching in the course of image acquisition will cause sophisticated deformations. Thus, in this study a hierarchical method is used to register images; the given method is the result of Rueckert et al.'s researches [34] and has been developed by Li et al. [18]. In this method global transformation and rotations are modeled using Affine transformation function and local deformations are corrected using free-form deformation model based on B-spline. Also, applying Jacobian operator to similarity measure makes no change in tumor volume during image registration process.

### 3.1.4 Extracting Region of Changes (EROC)

MRI artifacts can cause noise and forged changes which significantly influence the efficiency of presented methods as all parts of images need to be compared without considering important changes. In current section, we propose EROC solution to extract regions containing significant changes. In this method, at first, Mean Square Distance (MSD) function is used to compute similarity measure between two images, based on row by row and column by column intensity comparison and then similarity curves are plotted based on computed values. Fig. 4 shows an overview of EROC method. To better understand the proposed method, consider images A and B. MSD similarity measure can be computed for each row and column of these images through (2) to (5). In these equations, matrixes $MSD_T$, $MSD_D$, $MSD_R$, and $MSD_L$ are the results of top-down, bottom-up, right-to-left, and left-to-right traversals, respectively.

$$A : N \rightarrow M$$
$$B : N \rightarrow M$$

$$MSD_T(i) = (1/M) \sum_{K=1}^{M} (A(i,k) - B(i,k))^2$$ (2)

where $i = 1, 2, ...N$ and $MSD_T$ is a Matrix with size 1*N

$$MSD_{(i)} = MSD_T(N - i + 1)$$ (3)

where $i = 1, 2, ...N$ and $MSD_D$ is a Matrix with size 1*N

| | Histogram Image 2 | Histogram Image 1 | Image 2 (Follow-up 4) | Image 1 (Follow-up 1) |
|---|---|---|---|---|
| Before Normalization | 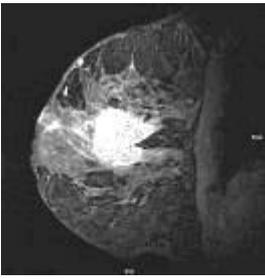 | 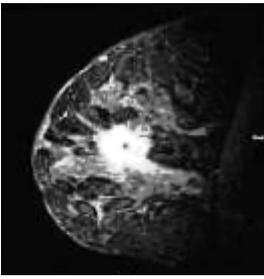 | 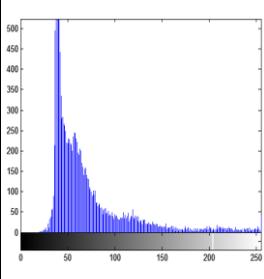 | 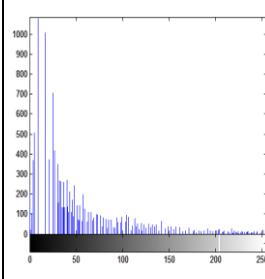 |
| After Normalization | 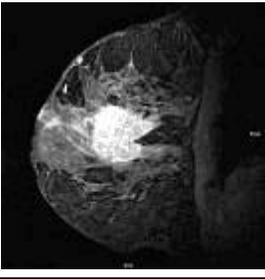 | 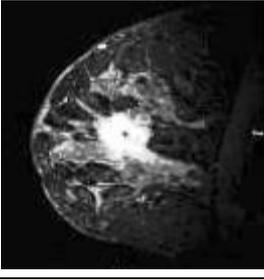 | 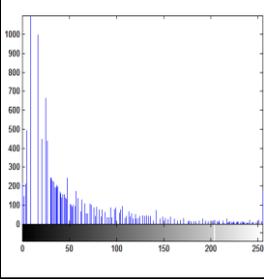 | 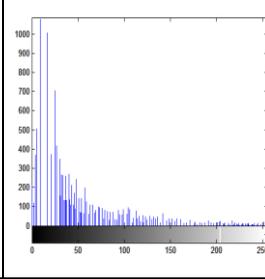 |

$$MSD_j = (1/N) \sum_{K=1}^{N} \left( A(k,j) - B(k,j) \right)^2 \quad (4)$$

where $j = 1, 2, \ldots M$ and $MSD_L$ is a Matrix with size 1*M

$$MSD_L(j) = MSD_L(M - j + 1) \quad (5)$$

where $j = 1, 2, \ldots M$ and $MSD_D$ is a Matrix with size 1*M

Once the similarity curves are plotted, they are given to Find First Point (FFP) algorithm to find first points of intensive changes. The flowchart in Fig. 5 shows the pseudo code, which is basically based on Hill-climbing algorithm with two major differences. First, in Hill-climbing algorithm the goal is finding the local maximum while in this function our purpose is finding the first ascending points to local maximum; second, local maximum points should be greater than threshold $T$ which is defined in (6).

$$T_{N,M} = (1/M) \sum_{c=1}^{K=N,M} \left( MSD[C]_{D,T \text{ if } R=N \text{ or } R,L \text{ if } R=M} \right) \quad (6)$$

where $K \in N$ and $K = N, M$

$T_M$ in (6) determines the required threshold value to find first points of changes during row by row traversal and this value can be computed using average value of matrix $MSD_T$ or $MSD_D$. $T_N$ is equal to $T_M$ during columnar traversal.

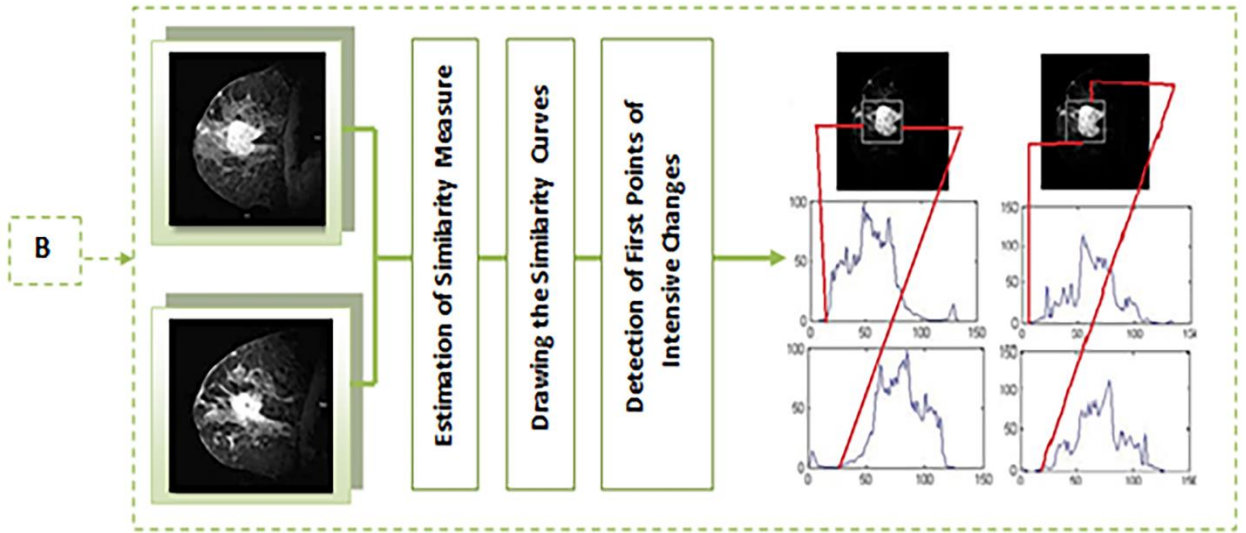

Fig. 4. An overview of the proposed EROC method.

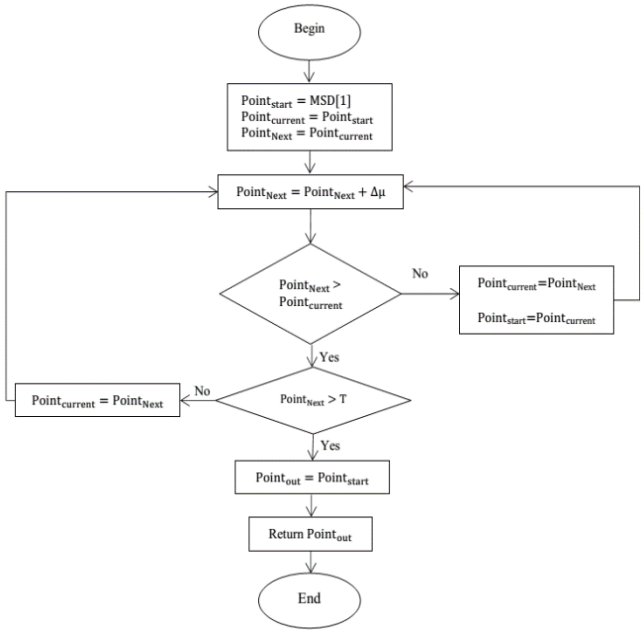

Fig. 5. FFPOC algorithm. A simple algorithm to find intensive change point.

In FFP algorithm, defining an optimum threshold to find first points of changes has a significant importance. The experimental results on both simulated and real images show that the estimated value of threshold *T* based on (6) is very close to optimum *T*. As it's shown in Fig. 6 smaller threshold value makes EROC method more sensitive to noise. As a result, the algorithm ends while achieving the first ascending points corresponding to the rows and columns containing pixels having noise. On the contrary, some significant changes will be lost on obtained region by higher threshold value.

## 3.2 Change Detection Phase

Direct intensity comparison is not an appropriate feature to detect changes in images as it's very sensitive to noise and requires a threshold to determine occurrence of a change. Since estimating exact value of this parameter requires prior knowledge about occurred changes, methods which are based on this feature won't have acceptable efficiency. Hence, in this study we use WLMI method to extract high level feature from each pixel to make better distinction between real and forged changes.

Also in contrast to previous methods, using an approach which is based on region growing can highlight the weak points of the thresholding method in defining pixels with significant changes. In fact, the main idea of this method is utilizing local consistency principle in both extracting WLMI features and adopting final decision about occurred changes since the significant ones occur locally and in interrelated pixels. Therefore, many pixels containing noise will be pruned and are not allowed to grow.

According to previous studies, in recent years the existing systems involve a great deal of preprocessing and also rely mostly on the statistical analysis such as GLRT [9]. In the following section, we introduce GLRT which is the most powerful presented method based on statistical models, and by addressing its weak points pave the way for a more detailed explanation of the proposed WLMI method.

### 3.2.1 GLRT

GLRT was first presented in 1998 by Kay [35]. In this method a fixed size N*N neighborhood $W$ is considered for each pixel and the ratio of the probabilities of two hy-

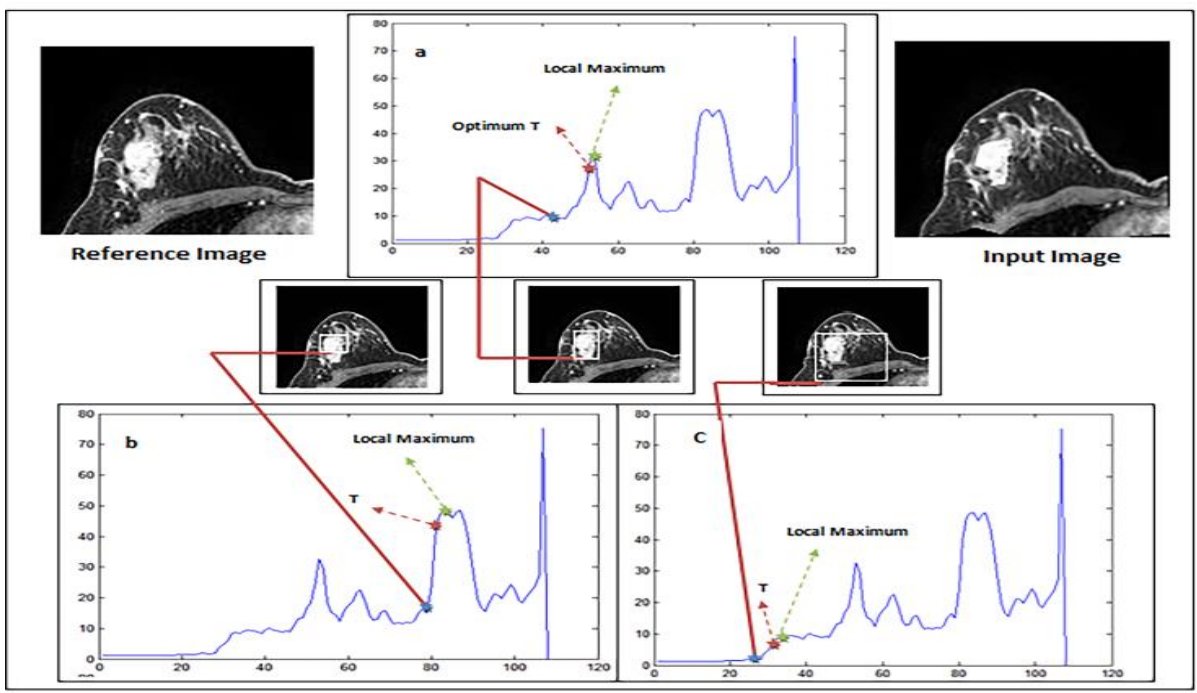

Fig. 6. Defining optimum threshold T.

potheses is computed based on neighborhood's statistical data [10]:

$H_0$: There is no change between $I_1$ and $I_2$ inside $W$.

$H_1$: There is a significant change between $I_1$ and $I_2$ inside $W$.

The probabilities are expressed in terms of parametrical probability distribution functions (pdfs). $H_0$ is interpreted using the pdf of $I_1$ and $I_2$ with the same parameter $\theta_0$, $H_1$ is interpreted using pdf of $I_1$ and $I_2$ with two different parameters $\theta_1 \neq \theta_2$. In general, GLRT can be computed as follows:

$$R_{GLRT} = \frac{\mathrm{P}(I_1; \theta_2)\,\mathrm{P}(I_2, \theta_2)}{\mathrm{P}(I_1; \theta_0)\,\mathrm{P}(I_2, \theta_0)}. \tag{7}$$

In conducted researches in medical field, the intensity values within $W$ are modeled as an average value plus Gaussian noise. Also, the noise variance is supposed constant throughout the image and is considered to be identical in both images and can be estimated by using the difference image. Therefore, the only pdf parameter $\theta$ that must be estimated is the average $\theta = \mu$ value. This value can be computed using (8):

$$\mu = \sum_{r \in w} I_r / n \tag{8}$$

, where $n$ is the number of pixels in $W$. Taking the logarithm of (8) gives the log-likelihood ratio:

$$l = \left(1/2\sigma^2\right) \sum_{p \in w} \begin{pmatrix} -\left(I_1(p) - \mu_1\right)^2 + \left(I_2(p) - \mu_2\right)^2 \\ + \left(I_1(p) - \mu_0\right)^2 + \left(I_2(p) - \mu_0\right)^2 \end{pmatrix}$$

which simplifies to:

$$\frac{\sqrt{n}}{2\sigma} |\mu_2 - \mu_1| \begin{array}{c} H_1 \\ > \\ < \\ H_0 \end{array} \gamma$$

$H_0$ is chosen if $l < \gamma$ where $\gamma$ is the threshold and can be computed by the ratio of probabilities of $H_0$ to $H_1$. Having no prior knowledge about how changes occur, can make estimating these values difficult, and influences the efficiency of the algorithm. The main weak points of this method are first, using average intensity as a comparison measure; second, assigning identical weight to each pixel in $W$ to determine the occurrence of changes in central pixel. In order to better demonstrate the weaknesses of this method, practical samples are shown in Fig. 7 and Fig. 8.

As it's shown in Fig. 7, no change has occurred in central pixel but GLRT method detects a pixel has changed. In this method due to use of average intensity, pixels containing intensive changes (such as noise or pixels adjacent to significant changes) influence the achieved result. Moreover, central pixel has the same role as the adjacent pixels in determining whether a change occurs or not. Now if we increase neighborhood size in order to eliminate the given weak points and to reduce effects of pixels containing noise, then subtle changes will not be recognizable. For example, Fig. 8 shows that a subtle change has occurred in the region of central pixel; however, GLRT method isn't able to recognize it. As neighborhood size becomes larger, average intensity measure acts like a smoothing filter and reduces the effect of changes of cen-tral pixel. Furthermore, identical weights are assigned to all pixels without considering their distance to central pixel.

### 3.2.2 Mutual Information Theory

Mutual information theory is widely used in image processing field as a measure to determine similarity between images [36]; therefore, in this research to dominate GLRT weak points, the idea of using this method to extract a feature with higher discrimination power than average intensity is discussed. In the following, we concisely express how to use this method in image processing field.

Consider an $n$ pixel image, the intensity values are the discrete values $X = \{X_1, X_2, \ldots, X_N\}$ where $X = 2^n$. According to the definitions, the entropy of intensity values in an image can be computed as follows:

$$H(X) = -\sum_{x \in X} p(x) \log p(x)$$

, where $p(x)$ represents probability distribution of each intensity value and can be achieved by histogram normalization.

Now suppose that the purpose is determining the amount of mutual information in a series of images. Following this assumption, consider two images to be the observations of two discrete random variables $X$ and $Y$ with probability distributions $p$ and $q$, respectively. The joint entropy $H(X, Y)$ for the discrete variables $X$ and $Y$ with joint probability distribution $r$ is defined as

$$H(X,Y) = -\sum_{x \in X} \sum_{y \in Y} r(x, y) \log r(x, y)$$

, which can be simplified to:

$$H(X,Y) = -\sum_{x,y} p(x)q(y) \log p(x) - \sum_{x,y} p(x)q(y) \log q(y)$$

$$= \mathrm{H}(X) + H(Y)$$

Mutual information theory can be defined with joint entropy equation as follows:

$$MI(X,Y) = \sum_{x \in X} \sum_{y \in Y} r(x, y) \log \left( r(x, y) / p(x) p(y) \right). \tag{9}$$

, and if assuming $X$ and $Y$ are independent then (9) can be simplified to (10) as follows:

$$MI(X,Y) = H(X) + H(Y) - H(X,Y). \tag{10}$$

### 3.2.3 Local Mutual Information

According to mutual information definition, it's obvious that this method is appropriate for large-scale issues and using it as a similarity measure in pixel level won't enjoy high efficiency. Thus, we require applying changes to (9), which can be usable in local condition. This work was done by Rogelj et al. [37] and is called local mutual information theory. Again consider (9), if we assume $i = [x,y]$ denotes the intensity pair of images $X$ and $Y$, then the equation can be rewritten in the following form.

$$MI(X,Y) = \sum_{i = [x,y]} p(i) \log \left( p(i) / \left( p(x) p(y) \right) \right) \tag{11}$$

If we do $p(i) = M_i / M$ replacement; where $M_i$ is the number of occurrences of intensity pair $i$, and $M$ is the total number of intensity pairs in the image, then (11) can be rewritten as (12).

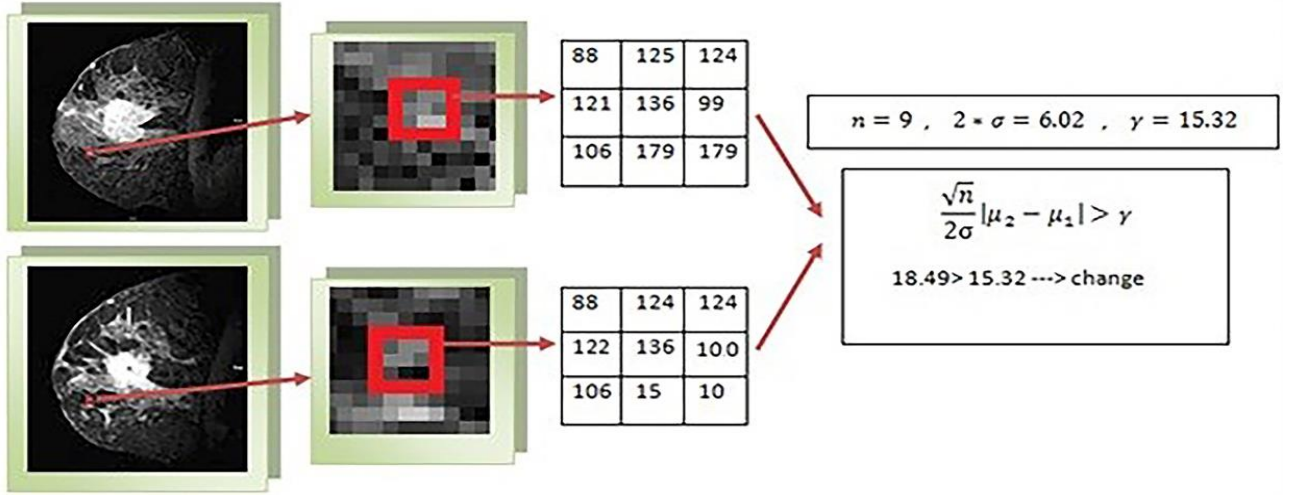

Fig. 7. A sample shows noise effect on GLRT efficiency with intensity changes.

$$MI(X,Y) = \sum_{i=[x,y]} (M_i/M) \log\left(p(i)\big/\left(p(x)\,p(y)\right)\right) \quad (12)$$

Let's consider neighborhood $W$ with size S*S. Based on above equation mutual information can be computed as shown in (13).

$$LMI(X,Y) = \sum_{i=[x,y]\in W_x} (M_i/M) \log\left(p(i)\big/\left(p(x)\,p(y)\right)\right) \quad (13)$$

, where $M_i$ and $M$ are defined the same as corresponding variables in (12), with an exception that these values are computed in region $W$. To better understand local mutual information theory, we can define above equations based on pixels in the given neighborhood as shown in (14).

$$LMI(X,Y) = \sum_{v\in W_x} (M_i/M) \log\left(p(i(v))\big/\left(p(x(v))\,p(y(v))\right)\right) \quad (14)$$

, where $v$ addresses pixels in neighborhood $W$.

In the following, we explain (14) which is used in our proposed method in order to compute similarity percentage of two pixels in serial images.

Let's consider two serial images $A(i,j)$ and $B(i,j)$; first, we define 3*3 neighborhood $W_{ij}$ for each pixel $X_{ij}$ in these images, then we compute similarity percentage of each $X_{ij}$ as follows:

$$SimRate_{ij} = LMI\left(W_{ij,A},W_{ij,B}\right)\big/ LMI\left(W_{ij,A},W_{ij,A}\right)*100 .$$

In this equation $LMI(W_{ij,A},W_{ij,B})$ indicates the maximum value of $LMI$ for neighborhood $W_{ij}$. In following section, the similarity percentage obtained by each pixel is used as a growing measure in region growing algorithm. In this method, as more number of pixels remain unchanged the similarity percentage gets increase. Also this method is not sensitive to intensive changes in adjacent pixels, since it works based on joint probability distribution. So, it's able to eliminate weak points of GLRT (i.e. see Fig. 7). In order to reduce this method's sensitivity to artifacts like illumination and global changes on intensity values in images; changes with $\Delta Intensity \leq 4$ will not be considered in joint histogram computation.

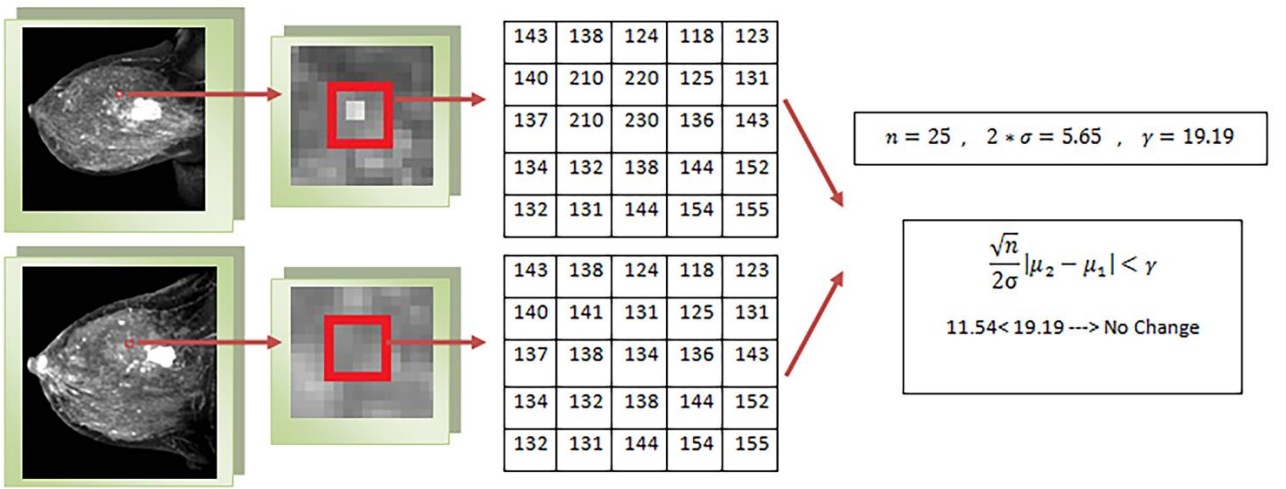

Fig. 8. A sample shows the effect of neighborhood's size and average measure on GLRT efficiency for subtle changes.

This method like GLRT will not be able to detect subtle changes in some cases due to assigning identical weight to all pixels. A sample of these subtle changes is shown in Fig. 9 which is not recognizable by LMI method. As we can see in this figure, significant changes have occurred in central pixels but since most of them remained unchanged, LMI method is not able to detect them, as identical weights assign to adjacent pixels. In the next section, we propose WLMI method in order to obviate weaknesses of GLRT and LMI methods.

### 3.2.4 Weighted Local Mutual Information

In the proposed WLMI method, pixels in neighborhood $W$ don't have same role in adopting final decision about occurrence of changes in central pixel, since different values have assigned to them.

We can compute WLMI as shown in (15).

$$WLMI(X,Y) =$$
$$\sum_{v \in Ws} (\beta_v M_i / M) \log\left( p(i(v)) \middle/ \left( p(x(v)) \right) p(y(v)) \right) \quad (15)$$

, where $\beta_v$ is the assigned weight to pixel $v$ in neighborhood $W_s$ and can be computed as follows:

$$\beta_v = \begin{cases} 2 \middle/ \left( N_{Change} * \exp\left( abs\left[ x(v) - y(v) \right] / 255 \right) \right) & \text{if } v = \text{center pixel} \\ 1 \middle/ \max\left[ \left( v_i^c - v_i \right), \left( v_j^c - v_j \right) \right] & \text{if } v \neq \text{center pixel} \end{cases}$$
(16)

In above equation, $N_{change}$ denotes pixels changed in neighborhood $W$, and $v^c$ is indicative of central pixels. According to (16), estimating central pixel's weight depends on two parameters $N_{change}$ and $\Delta I = x(v) - y(v)$. If no change occurs in central pixel, $\Delta I$ value will be equal to 1 and the only efficient parameter will be the number of changed pixels; and finally, the assigned weight to central pixel which is proportional to the number of unchanged pixels can result in increases of similarity percentage. But if central pixel changes, then the assigned weight will depend on intensity changes in addition to $N_{change}$ parameter; and also as $\Delta I$ value increases, similarity percentage reduces. In this method, assigned weights to adjacent pixels reduce as their distance to central pixel increase.

Similarity percentage of pixels in this method can be computed using LMI method as follows:

$$SimRate_{ij} = WLMI\left( W_{ij,A}, W_{ij,B} \right) \middle/ WLMI\left( W_{ij,A}, W_{ij,A} \right) * 100.$$

Extracted features by WLMI method can better differentiate between pixels on the basis of real and forged changes. To better understand this issue, Table 2 shows the comparison of features extracted by both the proposed method and GLRT. Image c is the difference image of given images, and significant changes are labeled in image e (see Table 2) by an expert. We can observe that the extracted features by GLRT method are not accurate to determine such changes.

Once again consider Fig. 7 and Fig. 10, to notice that the proposed method is able to detect this kind of changes in contrast to LMI and GLRT methods.

### 3.2.5 REGION GROWING METHOD

The final phase involves thresholding to adopt final decision about the occurrence of changes after determining similarity of extracted pixels. Estimating this threshold in change detection process is not practical due to required prior knowledge about the occurred changes. Therefore, to reduce the effect of thresholding in the efficiency of final system, we propose region-growing-based algorithm. Due to use of local consistency in this algorithm, many of forged changes are removed. First, pixels containing significant changes are selected as the seed points to grow, then the similarity between these pixels and adjacent pixels are computed, finally pixels which have the highest similarity to seed points are allowed to grow. This process will continue until no new pixel is found to be a growth candidate. In this theory several pixels containing noise are pruned and are not allowed to grow. As a result, the efficiency of algorithm increases in detecting significant changes. The proposed algorithm is shown in Table 3.

### Seed Selection

In this method, selecting seed points has significant importance. Since significant changes occurred around tumor, the distance from tumor position is an appropriate

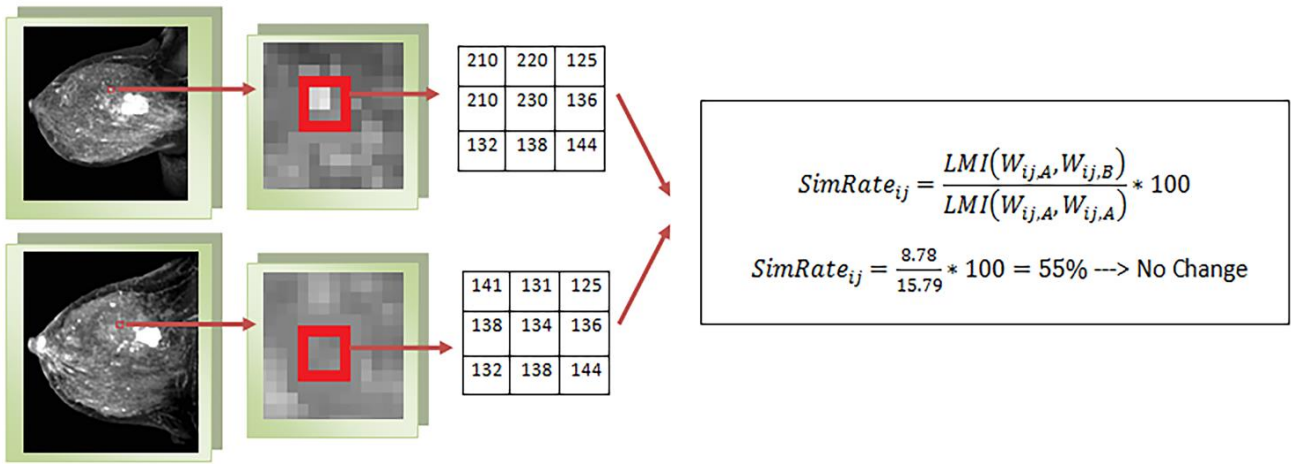

Fig. 9. A sample of subtle changes which LMI method is not able to recognize.

## TABLE 2
### THE COMPARISON OF EXTRACTED FEATURES BY THE PROPOSED METHOD AND GLRT METHOD

| Source Image(a) | Input Image(b) |
|---|---|
| 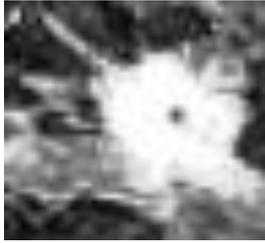 | 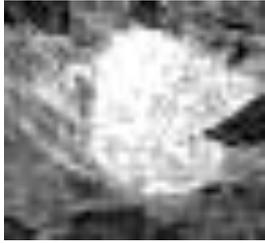 |

| Extracted Features by WLMI(f) | Extracted Features by GLRT(d) | Determined Changes by an Expert(e) | Difference Image(c) |
|---|---|---|---|
| 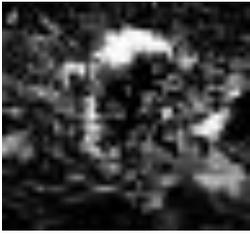 | 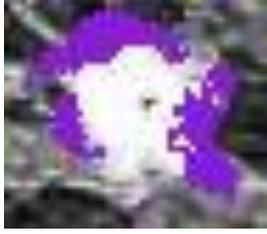 | 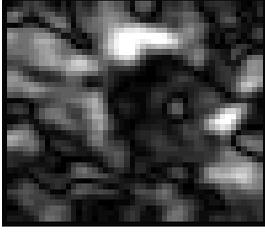 | 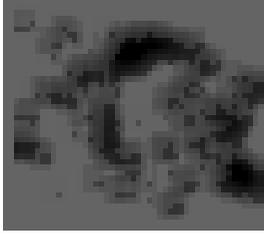 |

measure to select central points. Determining this distance requires tumor segmentation to identify its position in image. This action is possible through Improved Markov Random Field (I-MRF) method which we proposed in our previous researches [38]. After locating tumor, points with significant WLMI feature and a short distance from tumor, can be selected as seed points.

## Region Growing

In this phase, once again consider extracted features and central point $P_{seed}$. First, a fixed size 3*3 neighborhood $W_{ij}$ is determined for each central point, then each adjacent pixel is compared to central points and if the following constraint is satisfied then it's allowed to grow.

$SimRate_{ij}$ & $SimRate_{ij} < 50\%$

, where $SimRate'$ indicates similarity percentage of adja-

### TABLE 3
#### THE PROPOSED REGION GROWING ALGORITHM

**Seed Selection**
- Apply the I-MRF algorithm to extract tumor edge
- Determine the region of interest using extracted edge points
- In each row of these regions in difference image select a pixel with minimum WLMI

**Region Growing**
- calculate the WLMI similarity function for all neighbors of seeds and record them in a sorted list T in an Ascending order of similarity
- while T is not empty, remove the first point p and check its 8 neighbors. Add each of neighbors to T which they have a same similarity measure.

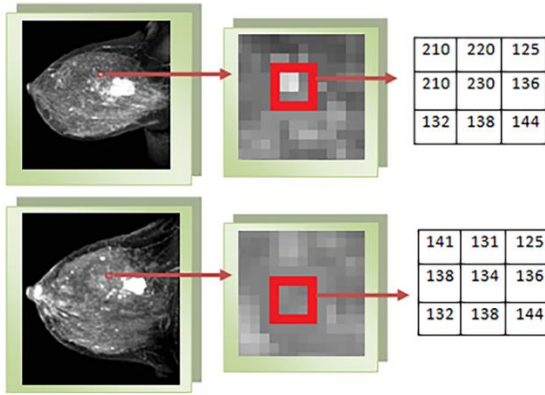

$$SimRate_{ij} = \frac{LMI(W_{ij,A}, W_{ij,B})}{LMI(W_{ij,A}, W_{ij,A})} * 100$$

$$SimRate_{ij} = \frac{8.78}{15.79} * 100 = 55\% \text{ ---> No Change}$$

$$SimRate_{ij} = \frac{WLMI(W_{ij,A}, W_{ij,B})}{WLMI(W_{ij,A}, W_{ij,A})} * 100$$

$$SimRate_{ij} = \frac{2.29}{5.87} * 100 = 39\% \text{ ---> Change}$$

Fig. 10. Eliminating weak points of LMI method with the proposed WLMI method.

cent pixels.

## 4 EVALUATION

The proposed change detection method has been evaluated by an expert on real images and simulated evolutions. In this section we describe our evaluation protocols.

Evaluating the results of presented methods in terms of change detection is complicated and also time-consuming, basically it stems from lack of ground truth for significant changes because generating such information is impossible for an expert due to following reasons: first, some changes are subtle and their intensities are close to other breast tissues therefore they are not recognizable by an expert; second, determining and labeling each changed pixel precisely is time-consuming, error prone and expensive, though we use the two following evaluation approaches:
- Using simulated images
- Evaluating achieved results on real images by an expert

Using simulated data provides the possibility of generating ground truth as it can be used to evaluate the effect of different parameters on performance of presented methods. In this research, simulated data is used with two main following purposes: 1) evaluating the power of presented method in detection of "evolving tumor" 2) evaluating system performance in detecting tumor evolution with different size in contrast to an expert's performance. As it is clear, simulated data will never have the complexity of real data therefore in this research another evaluation approach is used based on real images. Since generating ground truth for real data is impossible, first, images are given to system and then the achieved results are judged by an expert. Finally, aforementioned labels are used to compute precision measure.

### 4.1 Data Simulation

To create simulated images, PIDER breast MRI dataset [39] is used. This dataset contains MR images of 5 patients who suffer from breast cancer.

In order to create simulated data, in a way to be closer to real world issues, two fundamental features of real data must be considered:
1) Creating changes with different size in tumor volume
2) Creating complicated transformations which stem from patient position, image artifacts and etc. in real images.

### Creating Changes in Tumor Volume (Tumor Growth or Recovery)

We use presented method by Li et al. [18] to create changes in tumor volume. This algorithm is described in Table 3 to reduce tumor volume. Also simulation of tumor growth can be done in this way but the difference is that the first pixel is selected from an intact tissue. In this section, changes in different sizes from 5 to 90 percent of tumor volume are simulated in 27 images.

Tumor volume measure is an appropriate parameter to determine the rate of changes but let's consider a condition in which changes occur in some part of the tumor in an integrated manner, in this case the given changes can't be considered as small changes. Therefore, we use alternative measure called

TABLE 4
THE ALGORITHM OF SIMULATED DATA

- Segment the tumors manually to extract the pixels of Edge
- shrink tumors
    for every pixel of selected region to shrink tumor repeat:
    – select a pixel: P
    – calculate the distance between this pixel and all pixels of Edge
    – find a pixel with minimum distance: E
    – find mirror pixel $P_m$ in healthy tissue that satisfy $|PE| = |P_m E|$
    – smooth the intensity of $P_m$ by a 3*3 Gaussian filter
    – fill the pixel P with smoothed intensity

MD (Maximum Diameter) to determine this rate. By using this measure, better understanding of the rate of changes can be achieved. It's important to note that the size of each pixel is equal to 0.35 mm.

The aim of creating such changes is examination and evaluation of presented method in detecting changes with different size. Furthermore, subtle changes, around less than 10% of tumor volume are used to compare the performance of the proposed method with the ability of an expert in detecting such kind of changes.

TABLE 5
A SAMPLE OF SIMULATED IMAGES IN DIFFERENT SIZES

| | Tumor Recovery | Tumor Growth |
|---|---|---|
| Source Image | 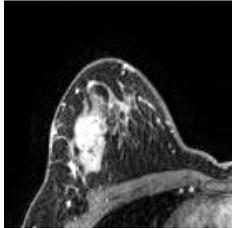 | 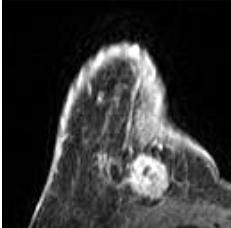 |
| Tumor With 30% Change | 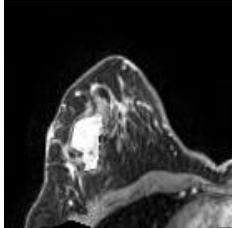 | 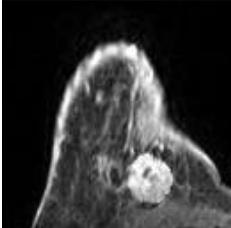 |
| Tumor With 70% Change | 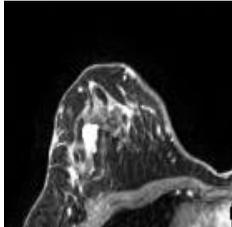 | 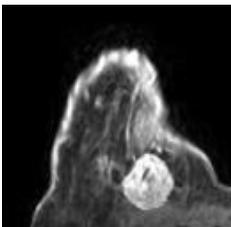 |

#### TABLE 6
##### A SAMPLE OF SIMULATED IMAGES CONTAINING SUBTLE AND UN-NOTICEABLE CHANGES

| Source Image | Tumor with 5% change |
|---|---|
| 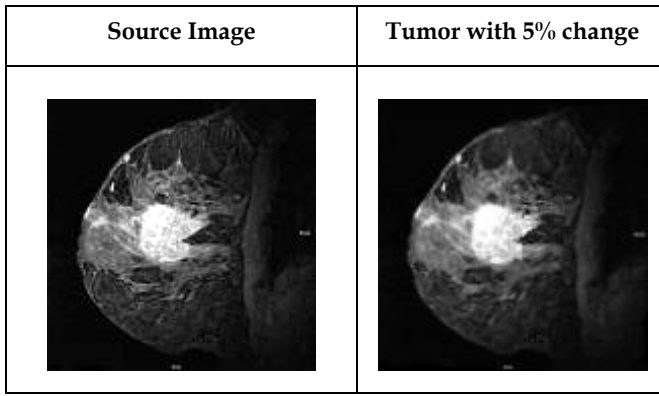 | |

#### TABLE 8
##### EVALUATION MEASURES

| Measure | Definition |
|---|---|
| Accuracy (ACC) | $(TP+TN)/(TP+FP+FN+TN)$ |
| Precision (PPV) | $TP/(TP+FP)$ |
| Specificity (SPC) | $TN/(TN+FP)$ |
| Sensitivity (TPR) | $TP/(TP+FN)$ |
| False Positive Rate (FPR) | $FP/(FP+TN)$ |
| Volume Overlap Ratio (VOR) | $\dfrac{p_c \cap p_s}{p_c \cup p_s} = \dfrac{TP}{TP+FP+FN}$ |

sizes. A sample of simulated images containing subtle and unnoticeable changes is shown in Table 6.

### 4.2 Evaluation Measures

Evaluation measures [2] used in this section are summarized in Table 7 and 8.

### 4.3 Experiment 1: Evaluating The Performance of the Proposed Framework On Simulated Images

In this experiment the performances of the proposed framework and GLRT method on simulated data are compared. In order to evaluate these methods in the same condition, GLRT algorithm is applied to regions extracted by EROC method.

Moreover, refer to [10] in order to implement and determine GLRT method parameters. The size of used neighborhood in this method is 3*3 and the noise variance is estimated by difference image.

The achieved results from experiment 1 are shown in Table 9. First column shows the overall comparison of both methods without considering the size of changes in tumor volume. A remarkable improvement is evident in PPV, VOR and ACC measures where in the proposed framework each one has increased 5.44%, 6% and 2.96% compared to GLRT method, respectively. By evaluating the effect of size of changes on the performances of given methods, results shown in second, third and fourth columns are accomplished. In the proposed method, the value of FPR is about 5% less than GLRT method in changes smaller than 10% of tumor volume. Another noticeable difference in PPV and ACC measures is that each of these measures is almost 10% and 3.49% higher than

### Creating Complicated Deformations

Creating images, which are closer to real data and contain all changes in reality, requires local and global deformation in addition to changes in size of tumors. To this aim, a mesh with the size of examined image is considered and then zero mean Gaussian and variance of three pixels are used to create distortion randomly. Finally, given mesh is applied to the image with image registration algorithm which is expressed in section 3.1.3. It is

#### TABLE 7
##### EVALUATION PARAMETERS

| | | Condition as Determined by "Simulator" | |
|---|---|---|---|
| | | Change | No-Change |
| **Test Outcome** | Change | True positive (TP) | False positive (FP) |
| | No-Change | False negative (FN) | True negative (TN) |

worth to note that for preserving the size of tumor, the given mesh is not applied to regions containing tumor. Table 5 shows a sample of simulated images in different

#### TABLE 9
##### COMPARING THE RESULT OF PROPOSED FRAMEWORK AND GLRT METHOD ON SIMULATED DATA WITH DIFFERENT SIZE

| | Overall Test | | Changes <10%*TV* MD <0.7mm | | 10%*TV* < Changes <30%*TV* 0.35mm < MD <1.05mm | | Changes >30%*TV* MD >1.05mm | |
|---|---|---|---|---|---|---|---|---|
| | Mean | | Mean | | Mean | | Mean | |
| | GLRT | WLMI | GLRT | WLMI | GLRT | WLMI | GLRT | WLMI |
| **VOR (%)** | 55.47 | 61.52 | 38.59 | 39.43 | 58.49 | 62.91 | 66.73 | 68.60 |
| **TPR (%)** | 83.69 | 89.54 | 66.99 | 68.25 | 94.14 | 93.27 | 89.94 | 93.97 |
| **FPR (%)** | 14.00 | 13.44 | 14.05 | 9.06 | 16.00 | 10.10 | 11.43 | 7.83 |
| **ACC (%)** | 86.11 | 89.07 | 86.21 | 89.70 | 86.20 | 89.86 | 87.48 | 92.97 |
| **SPC (%)** | 86.03 | 86.56 | 85.95 | 90.34 | 84.00 | 89.90 | 88.57 | 92.17 |
| **PPV (%)** | 60.81 | 66.25 | 41.02 | 51.23 | 60.96 | 67.59 | 75.31 | 72.52 |

*TV=Tumor Volume*

GLRT method, respectively. In general, it can be concluded that the precision of the proposed method is higher than GLRT method due to use of WLMI feature which can detect changes of this type.

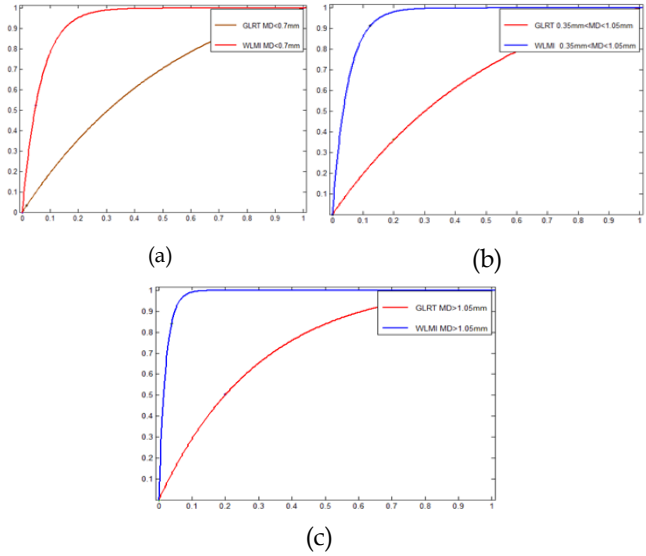

Fig. 11. ROC Curves (x axis: false positive rate, y axis: true positive rate), assessing the performance of proposed framework and GLRT in different size of simulated changes: (a) Changes <10%*TV*, (b) 10%*TV* < Changes <30%*TV*, (c) Changes>30%*TV*.

TABLE 10
THE VALUES OF A$_z$ FOR PROPOSED FRAMEWORK AND GLRT

| | AREA UNDER THE CURVE (A$_z$) | | |
|---|---|---|---|
| | CHANGES <10%TV | 10%TV < CHANGES <10%TV | CHANGES >30%TV |
| GLRT | 63.87 | 64.87 | 73.58 |
| WLMI | 93.49 | 94.97 | 97.92 |

The precedence of the proposed method in detecting changes for two other ranges is evident in the last two columns of Table 9, where the difference value of FPR and ACC measures for changes between 10% of tumor volume and 30% of tumor value is about 6.10% and 3.66%, respectively, and this value for changes greater than 30% of tumor value is about 3.6% and 5.49%, respectively.

To analyze the performance of each method more precisely, the ROC diagrams of accomplished results for all range of changes are drawn in Fig. 11 and the area under each curve are reported in Table 10.

Some samples of detected changes in simulated images are shown in Table 11 by using the proposed method and GLRT.

### 4.4 Experiment 2: Comparing the Performance of the Proposed Method with The Ability of an Expert in Detection of "Evolving Tumors"

The purpose of comparing the ability of an expert with the performance of the proposed method is evaluating their power on detecting subtle changes. In this experiment, 27 simulated images are shown to an expert and is asked to detect images containing changes (Whether a change has occurred in tumor-volume or not). In order to determine the ability of an expert, the ratio of the number of images containing changes, which an expert was able to detect, to total images is used as true positive measure. Since it's impossible for an expert to determine changes in pixel level, the evaluation measure is determined in image level. The achieved results are shown in Table 12.

TABLE 11
5 SAMPLES OF DETECTED CHANGES IN SIMULATED IMAGES USING PROPOSED METHOD AND GLRT

| | Test Image 1 | Test Image 2 | Test Image 3 | Test Image 4 | Test Image 5 |
|---|---|---|---|---|---|
| | Changes <10%TV | 10%TV<Changes<30%TV | | Changes>30%TV | |
| Ground Truth | | | | | |
| GLRT | | | | | |
| Proposed Framework | | | | | |

TABLE 12
COMPARING THE PERFORMANCE OF THE PROPOSED METHOD WITH THE ABILITY OF AN EXPERT

| | Human Expert (Image level) | | WLMI (Pixel level) | | | |
|---|---|---|---|---|---|---|
| | #Test Images | #TPs | #Test Images | #TPs | #Test Images | #TPs |
| Changes <10%TV | 17 | 3 | 17 | 3 | 17 | 3 |
| 10%TV<Changes<30%TV | 5 | 3 | 5 | 3 | 5 | 3 |
| Changes>30%TV | 5 | 5 | 5 | 5 | 5 | 5 |

As it's clear, the power of the proposed method in detecting subtle changes is higher than the expert's power since only 3 out of 17 images with changes less than 10% of tumor volume are detected by the expert. Also, it should be noted that evaluating the expert's ability in detecting changes greater than 30% of tumor volume doesn't indicate the weakness of the proposed method, as its true positive measure has been computed in image level.

## 4.5 Experiment 3: Evaluating the Performance of

TABLE 13
EVALUATING THE PERFORMANCE OF THE PROPOSED FRAMEWORK ON REAL IMAGES

| | WLMI | | | GLRT | | |
|---|---|---|---|---|---|---|
| | #TPs | #FPs | Precision (%) | #TPs | #FPs | Precision (%) |
| Test Image 1 | 272 | 140 | 66.01 | 274 | 79 | 77.62 |
| Test Image 2 | 246 | 118 | 67.58 | 178 | 116 | 60.54 |
| Test Image 3 | 43 | 26 | 62.31 | 0 | 4 | 0 |
| Test Image 4 | 372 | 136 | 73.22 | 101 | 59 | 63.12 |
| Test Image 5 | 265 | 44 | 85.76 | 188 | 151 | 55.45 |
| Test Image 6 | 58 | 24 | 70.73 | 76 | 155 | 32.90 |
| Test Image 7 | 381 | 281 | 57.55 | 188 | 190 | 49.73 |
| Test Image 8 | 196 | 38 | 83.76 | 103 | 61 | 62.80 |
| Test Image 9 | 428 | 316 | 57.52 | 164 | 152 | 51.89 |
| Test Image 10 | 152 | 38 | 80.00 | 110 | 44 | 71.42 |
| Test Image 11 | 173 | 18 | 90.57 | 375 | 244 | 60.58 |
| Test Image 12 | 152 | 154 | 49.67 | 193 | 255 | 43.08 |
| Test Image 13 | 991 | 363 | 73.19 | 543 | 314 | 63.36 |
| Test Image 14 | 1120 | 289 | 79.48 | 992 | 282 | 77.86 |
| Test Image 15 | 395 | 18 | 95.64 | 575 | 234 | 71.07 |
| Test Image 16 | 620 | 10 | 98.41 | 552 | 235 | 70.13 |
| Test Image 17 | 296 | 41 | 87.83 | 485 | 921 | 34.49 |
| Test Image 18 | 2875 | 1047 | 73.30 | 3581 | 2848 | 55.70 |
| Test Image 19 | 211 | 30 | 87.55 | 246 | 493 | 33.28 |
| Test Image 20 | 470 | 17 | 96.50 | 259 | 273 | 48.68 |
| Test Image 21 | 260 | 45 | 85.24 | 160 | 42 | 79.20 |
| Test Image 22 | 980 | 65 | 93.77 | 879 | 120 | 87.98 |
| Test Image 23 | 454 | 116 | 79.64 | 522 | 1210 | 30.13 |
| Test Image 24 | 233 | 377 | 38.19 | 270 | 630 | 30.00 |
| Test Image 25 | 1468 | 740 | 66.48 | 2150 | 1357 | 61.30 |
| Test Image 26 | 519 | 189 | 73.30 | 450 | 201 | 69.12 |
| Test Image 27 | 646 | 199 | 76.44 | 529 | 327 | 62.79 |
| Test Image 28 | 207 | 31 | 86.97 | 247 | 439 | 36.00 |
| Test Image 29 | 1610 | 467 | 77.51 | 647 | 264 | 71.02 |
| Test Image 30 | 1050 | 457 | 69.67 | 1771 | 508 | 77.70 |
| Max | 2875 | 1047 | 98.41 | 3581 | 2848 | 87.98 |
| Min | 43 | 10 | 38.20 | 0 | 4 | 0 |
| Average | 572 | 195 | 76.13 | 560.26 | 406.93 | 60.27 |

**the Proposed Framework on Real Images**

In this section, 30 real MR images related to patients suffering from breast cancer are collected in different cycles of chemotherapy from Imam Reza Hospital [40] and internet references.

Since there is no possibility to generate ground truth for these images, the final evaluation is conducted based on the expert's opinion. The evaluation's protocol constitutes the three following steps:

1. Detecting changes in tumor volume by proposed method
2. Labeling detected changes of previous step by an expert in two groups: TPs (highlighted pixels in green color), FPs (highlighted pixels in red color).
3. Computing the precision of the proposed method

The results of the proposed method and GLRT method on real images are reported in Table 13. On average, the precision of the proposed method in extracting changes from real images is 16% higher than GLRT method in the same condition. Also one real sample of detected changes using the above described protocol is shown in Table 14.

# 6 DISCUSSION AND FUTURE WORK

As it was shown in experimental results, many of the occurred changes in tumors were not recognizable by the manual change detection. Exploring all parts of two images manually to detect change is a very hard and error prone task. Therefore, experts tend to skip areas that may contain valid evolutions. Furthermore, the expert can adopt wrong decision based on acquired artifacts. As an example, contrast change causes tumor-volume to appear larger or smaller than its real size. The proposed framework in this research can solve this issue to some extent and a tool is provided that can help an expert to find lesion evolutions more accurately

Comparing all parts of two images is time-consuming and can reduce efficiency since the locations of the occurred changes are not considered. In proposed framework, using EROC method to extract regions with significant changes can increase the precision of system (PPV measure) about 21%. The precision of the system with and without use of EROC method is reported in Table 15. It must be noted that change detection process in medical images faces with risk of losing significant changes during removing artifacts. Consider the first row of the Table 15, system's TPR is reduced about 0.5% due to use of EROC method. In this system, the negligible error caused by EROC method can be ignored for 21% improvement in precision measure.

Another noticeable feature in this system is the time required to check images compared to amount of time used by the expert. According to the achieved results in the last row of Table 15, the proposed method scans 27 images in less than 81(27 * 3) seconds due to use of EROC algorithm, while it takes 15 minutes for an expert to compare two images.

Finally, using the proposed WLMI method in contrast to GLRT to extract high level features results in 5.44% and 15.85% detection power improvement, respectively, in simulated and real image set.

As one of the weak points of the proposed system, we can address to experimental estimation of the threshold value in EROC method based on the prior knowledge existed in images. Furthermore, when EROC method faces with newfound tumors, it doesn't enjoy an acceptable efficiency. As an example, the newfound tumor shown in Table 16 in red circle causes the extracted region to contain some unreal changes. To eliminate this issue, we intend to use Demons algorithm [41] to locally extract regions containing changes.

The proposed framework we have described here is a general change detection system and it is able to detect changes in benign and malignant tumors in other MRI images that are obtained from other organs. But it is completely inefficient in detection of changes in some cases such as Microcalcifications.

As we know, using unsupervised methods is not a good choice in issues which have high complexity.

TABLE 14
ONE REAL SAMPLE OF DETECTED CHANGES IN REAL IMAGES
USING THE PROPOSED METHOD AND GLRT

TABLE 15
THE PRECISION AND COMPUTATIONAL TIME OF THE PROPOSED FRAMEWORK WITH AND WITHOUT USE OF EROC METHOD

|  | WLMI(Without EROC) | | | WLMI(With EROC) | | |
|---|---|---|---|---|---|---|
|  | **Mean** | **Max** | **Min** | **Mean** | **Max** | **Min** |
| **TPR (%)** | 90.01 | 98.99 | 61.61 | 89.54 | 99.24 | 61.61 |
| **PPV (%)** | 44.35 | 67.81 | 17.52 | 66.25 | 76.74 | 41.82 |
| **Computational Time (s)** | 29.56 | 35.67 | 25.23 | 3.16 | 5.07 | 0.76 |

Though, using supervised methods and collecting prior knowledge about the issues can diminish the complexity of them to some extent. Hence, in future we intend to use model-based methods to extract regions and to detect changes in microcalcifications, since these anomalies in breast are very similar to noises in the image and without extracting distinctive features we won't be able to detect them.

TABLE 16
AN EXAMPLE OF NEWFOUND TUMOR IN BREAST MR IMAGES

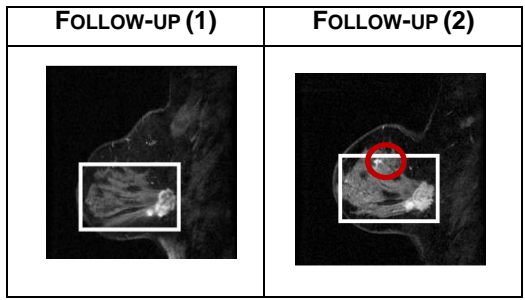

| FOLLOW-UP (1) | FOLLOW-UP (2) |
|---|---|

In the end, as it was mentioned in the paper's title, the proposed method is fully automatic but it's clear that the final decision should be adopted by an expert about detected changes. In fact, the final purpose of creating this system is producing a tool which can help doctors to detect changes more accurately.

## 7 CONCLUSION

In this research, we propose a comprehensive framework to automatically detect changes in serial breast MR images, even the subtle ones which are missed by an expert. In preprocessing phase, by extracting regions containing significant changes, many forge and unimportant changes are removed in the beginning and thereby the final efficiency increases. In change detection phase, in order to eliminate weak points of previous methods, WLMI method is proposed to extract high level features. Given features in comparison to other measures such as intensity values can better differentiate between forged and real images. In final phase, region growing algorithm is proposed to adopt final decision about the occurrence of changes, and as this algorithm follows local consistency principle, pixels containing noise are not allowed to grow. Therefore, false positive rate of the proposed framework is noticeably reduced. Obtained results from evaluation of the presented method in real and simulated images show the precedence of the proposed method compared to statistic methods like GLRT.

## REFERENCES

[1] E.D. Pisano, C. Gatsonis, E. Hendrick, M. Yaffe, J.K. Baum, S. Acharyya, E.F. Conant, L.L. Fajardo, L. Bassett, C. D'Orsi, R. Jong, M. Rebner, "Diagnostic Performance of Digital versus Film Mammography for Breast-Cancer Screening," *N Engl J Med*, vol. 353, no. 17, pp. 1773-1783, Oct. 2005.

[2] R.J. Smith, V. Cokkinides, H.J. Eyre, "Cancer Screening in the United States, A Review of Current Guidelines, Practices, and Prospects," CA Cancer J Clin, vol. 57, no. 2, pp. 90-104, 2007.

[3] D.B. Kopans, "The Positive Predictive Value of Mammography," *AJR Am J Roentgenol*, vol. 158, no. 3, pp. 521-526, Mar. 1992.

[4] E.A. Morris, "Diagnostic Breast MR Imaging: Current Status and Future Directions," *Radiol Clin North Am*, vol. 45, no. 5, pp. 863–880, 2007.

[5] M. Abedi, D. Farrokh, F. ShandizHomaei, A. Joulaee, R. Anbiaee, B. Zandi, "The Validity of MRI in Evaluation of Tumor Response to Neoadjuvant Chemotherapy in Locally Advanced Breast Cancer," *Iran J Cancer Prev*, 6(1), pp. 28–35, 2013.

[6] J.H. Chen, B.A. Feig, D.J. Hsiang, J.A. Butler, R.S. Mehta, S. Bahri, O. Nalcioglu, M.Y. Su, "Impact of MRI-Evaluated Neoadjuvant Chemotherapy Response On Change of Surgical Recommendation in Breast Cancer," *Ann Surg*, vol. 249, no. 3, pp. 448-454, 2009.

[7] E.D. Angelini, J. Delon, A.B. Bah, L. Capelle, E. Mandonnet, "Differential MRI Analysis For Quantification Of Low," *Med Image Anal* , vol. 16, no. 1, pp. 114-126, Jan. 2012.

[8] M. Battaglini, F. Rossi, R.A. Grove, M.L. Stromillo, B. Whitcher, P.M. Matthews, N. De Stefano, "Automated Identification of Brain New Lesions in Multiple Sclerosis Using Subtraction Images," *J Magn Reson Imaging*, vol. 39, no. 6, pp. 1543-1549, June 2014.

[9] H. Boisgontier, N. Vincent, F. heitz, L. Rumbach, J.P. Armspach, "Generalized Likelihood Ratio Tests for Change Detection in Diffusion Tensor Images: Application to Multiple Sclerosis," *Med Image Anal*, vol. 16, no. 1, pp. 325-338, Jan. 2012.

[10] M. Bosc, F. Heitz, J.P. Armspach, I. Namer, D. Gounot, L. Rumbach, "Automatic Change Detection in Multimodal Serial MRI: Application to Multiple Sclerosis Lesion Evolution," *NeuroImage*, vol. 20, no. 2, pp. 643–656, Oct. 2003.

[11] O. Ganiler, A. Oliver, Y. Diez, J. Freixenet, J.C. Vilanova, B. Beltran, L. Ramió-Torrentà, A. Rovira, X. Lladó, "A Subtraction Pipeline for Automatic Detection of New Appearing Multiple Sclerosis Lesions in Longitudinal Studies," *Neuroradiology*. vol. 56, no. 5, pp. 363-374, May 2014.

[12] M.A. Horsfield, M.A. Rocca, E. Pagani, L. Storelli, P. Preziosa, R. Messina, F. Camesasca, M. Copetti, M. Filippi, "Estimating Brain Lesion Volume Change in Multiple Sclerosis by Subtraction of Magnetic Resonance Images," *Neuroimaging*, vol. 26, no. 4, pp. 395-402, Jul. 2016.

[13] V. Nika, P. Babyn, H. Zhu, "Change Detection of Medical Images Using Dictionary Learning Techniques and Principal Component Analysis," *J Med Imaging*, vol. 1, no. 2, Jul. 2014.

[14] J. Patriarche, B. Erickson, "Part 1. Automated Change Detection and Characterization in Serial MR Studies of Brain-Tumor Patients," *J. Digital Imaging*, vol. 20, no. 3, pp. 203-222, Sep. 2007.

[15] J. Patriarche, B. Erickson, "Part 2. Automated Change Detection and Characterization Applied to Serial MR of Brain Tumors May Detect Progression Earlier Than Human Experts," *J. Digital Imaging*, vol. 20, no. 4, pp. 321-328.

[16] H. Seo, P. Milanfar, "A Non-Parametric Approach to Automatic Change Detection in MRI Images of the Brain," *Proc. IEEE Symp. Biomedical Imaging: From Nano to Macro (ISBI '09)*, pp. 245-248, 2009, DOI: 10.1109/ISBI.2009.5193029.

[17] N.C. Atuegwu, X. Li, L.R. Arlinghaus, R.G. Abramson, J.M. Williams, A.B. Chakravarthy, V.G. Abramson, T.E. Yankeelov, "Longitudinal, Intermodality Registration of Quantitative Breast PET and MRI Data Acquired Before and During Neoadjuvant Chemotherapy: Preliminary Results," *Med Phys*, vol. 41, no. 5, May 2014, DOI: 10.1118/1.4870966.

[18] X. Li, B.M. Dawant, E.B. Welch, A.B. Chakravarthy, D. Freehardt, I. Mayer, M. Kelley, I. Meszoely, J.C. Gore, T.E. Yankeelov, "A Non-Rigid Registration Algorithm for Longitudinal Breast MR Images and the Analysis of Breast Tumor Response," *Magn Reson Imaging*, vol. 27, no. 9, pp. 1258-1270, Nov. 2009.

[19] X. Li, L.R. Arlinghaus, A. Chakravarthy, J. Farley, I. Mayer, V. Abramson, M. Kelley, I. Meszoely, J. Means-Powell, T.E. Yankeelov, "Early DCE-MRI changes after longitudinal registration may predict breast cancer response to Neoadjuvant Chemotherapy," Proceedings of the 5th international conference on Biomedical Image Registration, pp. 229-235, July 2012.

[20] Y. Ou, S.P. Weinstein, E.F. Conant, S. Englander, X. Da, B. Gaonkar, M.K. Hsieh, M. Rosen, A. DeMichele, C. Davatzikos, D. Kontos, "Deformable Registration for Quantifying Longitudinal Tumor Changes During Neoadjuvant Chemotherapy," *Magn Reson Med*, vol. 73, no. 6, pp. 2343-2356, June 2015.

[21] J. Wu, Y. Ou, S.P. Weinstein, E.F. Conant, N. Yu, V. Hoshmand, B. Keller, A.B. Ashraf, M. Rosen, A. DeMichele, C. Davatzikos, D. Kontos, "Quantification of Tumor Changes During Neoadjuvant Chemotherapy with Longitudinal Breast DCE-MRI Registration," *Proc. SPIE Medical Imaging Symp.*, Mar. 2015, doi: 10.1117/12.2081938.

[22] F. Aghaei, M. Tan, A.B. Hollingsworth, B. Zheng, "Applying A New Quantitative Global Breast MRI Feature Analysis Scheme To Assess Tumor Response To Chemotherapy," *Journal of Magnetic Resonance Imaging*, vol. 44, no. 5, pp. 1099-1106, Nov. 2016.

[23] C.P. Chou, M.T. Wu, H.T. Chang, Y.S. Lo, H.B. Pan, H. Degani, H. Furman-Haran "Monitoring Breast Cancer Response to Neoadjuvant Systemic Chemotherapy Using Parametric Contrast-Enhanced MRI: A Pilot Study," *Acad Radiol*, vol. 14, no. 5, pp. 561–73, May 2007.

[24] N. Michoux, S. Van den Broeck, L. Lacoste, L. Fellah, C. Galant, M. Berlière, I. Leconte, "Texture Analysis on MR Images Helps Predicting Non-Response to Nac in Breast Cancer," *BMC Cancer*, vol. 15, no. 574, Aug. 2015, DOI: 10.1186/s12885-015-1563-8.

[25] A. Oliver, M. Tortajada, X. Lladó, J. Freixenet, S. Ganau, L. Tortajada, M. Vilagran, M. Sentís, R. Martí, "Breast Density Analysis Using an Automatic Density Segmentation Algorithm," *Journal of Digital Imaging*, vol. 28, no. 5, pp. 604-612, Oct. 2015.

[26] K. Nie, J. Chen, S. Chan, M. Chau, H. Yu, S. Bahri, T. Tseng, O. Nalcioglu, M. Su, "Development of A Quantitative Method for Analysis of breast density based on Three-Dimensional Breast MRI," *Med Phys*, vol. 35, no. 12, pp. 5253-5262, Dec. 2008.

[27] S. Partridge, J. Gibbs, Y. Lu, L. Esserman, D. Tripathy, D. Wolverton, H. Rugo, E. Hwang, C. Ewing, N. Hylton, "MRI measurements of Breast Tumor Volume Predict Response to neoadjuvant chemotherapy and Recurrence-Free Survival," *Am J Roentgenol*, vol. 184, no. 6, pp. 1774-1781, June 2005.

[28] S. Smith, N. De Stefano, M. Jenkinson, P. Matthews, "Measurement of Brain Change Over Time," *FMRIB Technical Report TR00SMS1*, 2000.

[29] S. Yousefi, R. Azmi, M. Zahedi, "Brain Tissue Segmentation in MR Images Based on A Hybrid of MRF and Social Algorithms," *Med Image Analysis*, vol. 16, no. 4, pp. 840–848, May 2012.

[30] R. Azmi, B. Pishgoo, N. Norozi, S. Yeganeh, "Ensemble Semi-Supervised Frame-Work for Brain Magnetic Resonance Imaging Tissue Segmentation," *J. Med. Signals. Sens*, vol. 3, no. 2, pp. 94–106, 2013.

[31] R. Azmi, N. Norozi, L. Salehi, A. Amirzadi, R. Anbiaee, "IMPST: A New Interactive Self-Training Approach to Segmentation Suspicious Lesions in Breast MRI," *J Med Signals Sens*, vol. 1, no. 2, pp. 138–148, 2011.

[32] L. Lemieux, U. Wieshmann, N. Moran, D. Fish, "The Detection and Significance of Subtle Changes in Mixed-Signal Brain Lesions by Serial MRI Scan Matching and Spatial Normalization," *Med. Image Anal*, vol. 2, no. 3, pp. 227-242, Sep. 1998.

[33] R.D. Nowak, "Wavelet-Based Rician Noise Removal for Magnetic Resonance Imaging," *IEEE Trans. on Image Processing*. vol. 8, no. 10, pp. 1408-1419, 1999, DOI: 10.1109/83.791966.

[34] D. Rueckert, L.I. Sonoda, C. Hayes, D.L.G. Hill, M.O. Leachand, D.J. Hawkes, "Non-rigid Registration Using Free-Form Deformations: Application to Breast MR Images," *IEEE Trans. Med. Imaging*, vol. 18, no. 8, pp. 712– 721.

[35] S. Kay, "Fundamentals of statistical signal processing: detection theory," Prentice Hall, 1993.

[36] J.P.W. Pluim, J.B.A. Maintz, M.A. Viergever, "Mutual-Information Based Registration of Medical Images: A Survey," *IEEE Trans. Med. Imag.* vol. 22, no. 8, pp. 986–1004.

[37] P. Rogelj, S. Kovacic, "Point Similarity Measure Based on Mutual Information," *Biomedical Image Registration*, J.C. Gee, A. Maintz, M.W. Vannier, Lecture Notes in Computer Science2717, Berlin: Springer-Verlag, pp. 112–121, 2003.

[38] R. Azmi, N. Norouzi, "A New Markov Random Field Segmentation Method for Breast Lesion Segmentation in MR Images," *J Med Signals Sens*, vol. 1, no. 3, pp. 156-164, 2011.

[39] "NBIA-National Biomedical Imaging Archive," 2016; https://imaging.nci.nih.gov/ncia.

[40] "Mashhad University of Medical Sciences," 2016; http://www.mums.ac.ir/newmums/index.php/en.

[41] H. Wang, L. Dong, J. O'Daniel, R. Mohan, A.S. Garden, K.K. Ang, D.A. Kuban, M. Bonnen, J.Y. Chang, R. Cheung, "Validation of an Accelerated 'Demons' Algorithm for Deformable Image Registration in Radiation Therapy," *Phys. Med. Biol.* vol. 50, no. 12, pp. 2887–2905, June 2005.